\documentclass[prd,aps,showpacs,preprintnumbers,amsmath,amssymb,superscriptaddress]{revtex4}
\usepackage{alltt}
\usepackage{epsfig}
\usepackage{calc,rotating,xspace}
\usepackage{pifont}

\newcommand{\Ds}{\ensuremath{D_s^+} }
\newcommand{\Dd}{\ensuremath{D^+} }

\newcommand{\DsNoSpc}{\ensuremath{D_s^+}}

\newcommand{\GeV}{\ensuremath{\mathrm{Ge\kern -0.1em V}}}
\newcommand{\MeV
}{\ensuremath{\mathrm{Me\kern -0.1em V}}}

\newcommand{\um}{\ensuremath{\mathrm{\mu m}}}

\newcommand{\MeVCSq}{\ensuremath{\MeV\!/c^2}}
\newcommand{\GeVCSq}{\ensuremath{\GeV\!/c^2}}
\newcommand{\GeVC}{\ensuremath{\GeV\!/c}}

\begin{document}
\clearpage


\title{Measurement of the Mass Difference 
       \boldmath$ m(\Ds)-m(\Dd)$ \unboldmath at CDF II}


\affiliation{Institute of Physics, Academia Sinica, Taipei, Taiwan 11529, Republic of China }
\affiliation{Argonne National Laboratory, Argonne, Illinois 60439 }
\affiliation{Istituto Nazionale di Fisica Nucleare, University of Bologna, I-40127 Bologna, Italy }
\affiliation{Brandeis University, Waltham, Massachusetts 02254 }
\affiliation{University of California at Davis, Davis, California 95616 }
\affiliation{University of California at Los Angeles, Los Angeles, California 90024 }
\affiliation{University of California at Santa Barbara, Santa Barbara, California 93106 }
\affiliation{Instituto de Fisica de Cantabria, CSIC-University of Cantabria, 39005 Santander, Spain }
\affiliation{Carnegie Mellon University, Pittsburgh, Pennsylvania 15213 }
\affiliation{Enrico Fermi Institute, University of Chicago, Chicago, Illinois 60637 }
\affiliation{Joint Institute for Nuclear Research, RU-141980 Dubna, Russia }
\affiliation{Duke University, Durham, North Carolina 27708 }
\affiliation{Fermi National Accelerator Laboratory, Batavia, Illinois 60510 }
\affiliation{University of Florida, Gainesville, Florida 32611 }
\affiliation{Laboratori Nazionali di Frascati, Istituto Nazionale di Fisica Nucleare, I-00044 Frascati, Italy }
\affiliation{University of Geneva, CH-1211 Geneva 4, Switzerland }
\affiliation{Glasgow University, Glasgow G12 8QQ, United Kingdom }
\affiliation{Harvard University, Cambridge, Massachusetts 02138 }
\affiliation{University of Helsinki, FIN-00044, Helsinki, Finland }
\affiliation{Hiroshima University, Higashi-Hiroshima 724, Japan }
\affiliation{University of Illinois, Urbana, Illinois 61801 }
\affiliation{The Johns Hopkins University, Baltimore, Maryland 21218 }
\affiliation{Institut f\"ur Experimentelle Kernphysik, Universit\"at Karlsruhe, 76128 Karlsruhe, Germany }
\affiliation{High Energy Accelerator Research Organization (KEK), Tsukuba, Ibaraki 305, Japan }
\affiliation{Center for High Energy Physics: Kyungpook National University, Taegu 702-701; Seoul National University, Seoul 151-742; and SungKyunKwan University, Suwon 440-746; Korea }
\affiliation{Ernest Orlando Lawrence Berkeley National Laboratory, Berkeley, California 94720 }
\affiliation{University of Liverpool, Liverpool L69 7ZE, United Kingdom }
\affiliation{University College London, London WC1E 6BT, United Kingdom }
\affiliation{Massachusetts Institute of Technology, Cambridge, Massachusetts 02139 }
\affiliation{University of Michigan, Ann Arbor, Michigan 48109 }
\affiliation{Michigan State University, East Lansing, Michigan 48824 }
\affiliation{Institution for Theoretical and Experimental Physics, ITEP, Moscow 117259, Russia }
\affiliation{University of New Mexico, Albuquerque, New Mexico 87131 }
\affiliation{Northwestern University, Evanston, Illinois 60208 }
\affiliation{The Ohio State University, Columbus, Ohio 43210 }
\affiliation{Okayama University, Okayama 700-8530, Japan }
\affiliation{Osaka City University, Osaka 588, Japan }
\affiliation{University of Oxford, Oxford OX1 3RH, United Kingdom }
\affiliation{Universit\'a di Padova, Istituto Nazionale di Fisica Nucleare, Sezione di Padova-Trento, I-35131 Padova, Italy }
\affiliation{University of Pennsylvania, Philadelphia, Pennsylvania 19104 }
\affiliation{Istituto Nazionale di Fisica Nucleare, University and Scuola Normale Superiore of Pisa, I-56100 Pisa, Italy }
\affiliation{University of Pittsburgh, Pittsburgh, Pennsylvania 15260 }
\affiliation{Purdue University, West Lafayette, Indiana 47907 }
\affiliation{University of Rochester, Rochester, New York 14627 }
\affiliation{Rockefeller University, New York, New York 10021 }
\affiliation{Instituto Nazionale de Fisica Nucleare, Sezione di Roma, University di Roma I, ``La Sapienza," I-00185 Roma, Italy }
\affiliation{Rutgers University, Piscataway, New Jersey 08855 }
\affiliation{Texas A\&M University, College Station, Texas 77843 }
\affiliation{Texas Tech University, Lubbock, Texas 79409 }
\affiliation{Institute of Particle Physics, University of Toronto, Toronto M5S 1A7, Canada }
\affiliation{Istituto Nazionale di Fisica Nucleare, University of Trieste/\ Udine, Italy }
\affiliation{University of Tsukuba, Tsukuba, Ibaraki 305, Japan }
\affiliation{Tufts University, Medford, Massachusetts 02155 }
\affiliation{Waseda University, Tokyo 169, Japan }
\affiliation{University of Wisconsin, Madison, Wisconsin 53706 }
\affiliation{Yale University, New Haven, Connecticut 06520 }


\author{D.~Acosta}
\affiliation{University of Florida, Gainesville, Florida 32611 }

\author{T.~Affolder}
\affiliation{University of California at Santa Barbara, Santa Barbara, California 93106 }

\author{M.~H.~Ahn}
\affiliation{Center for High Energy Physics: Kyungpook National University, Taegu 702-701; Seoul National University, Seoul 151-742; and SungKyunKwan University, Suwon 440-746; Korea }

\author{T.~Akimoto}
\affiliation{University of Tsukuba, Tsukuba, Ibaraki 305, Japan }

\author{M.~G.~Albrow}
\affiliation{Fermi National Accelerator Laboratory, Batavia, Illinois 60510 }

\author{B.~Alcorn}
\affiliation{Fermi National Accelerator Laboratory, Batavia, Illinois 60510 }

\author{C.~Alexander}
\affiliation{University of Pennsylvania, Philadelphia, Pennsylvania 19104 }

\author{D.~Allen}
\affiliation{Fermi National Accelerator Laboratory, Batavia, Illinois 60510 }

\author{D.~Allspach}
\affiliation{Fermi National Accelerator Laboratory, Batavia, Illinois 60510 }

\author{P.~Amaral}
\affiliation{Enrico Fermi Institute, University of Chicago, Chicago, Illinois 60637 }

% ...

\collaboration{ CDF II Collaboration }
\noaffiliation

\pacs{13.25.Ft, 14.40.Lb}

\begin{abstract}
  \noindent
  We present a measurement of the mass difference $m(\Ds)-m(\Dd)$, where both
  the \Ds and \Dd are reconstructed in the $\phi \pi^+$ decay channel. This
  measurement uses $11.6~\rm{pb^{-1}}$ of data collected by CDF II using the
  new displaced-track trigger. The mass difference is found to be
  $$
  m(\Ds)-m(\Dd) = \rm 99.41 \pm 0.38(stat.) \pm 0.21(syst.)~\MeVCSq.
  $$

\end{abstract}
\maketitle

\section{Introduction}

Meson masses are predicted by different models of quark interactions and the
inter-quark potential.
Analytically, the spectrum of heavy-light mesons can be described in the
QCD framework using the principles of Heavy Quark Symmetry and Heavy Quark
Effective Theory \cite{mass-analytical, mass-inbook}.
These theories state that in the limit of infinitely heavy quark mass, the
properties of the meson are independent of the heavy quark flavor and that the
heavy quark does not contribute to the orbital degrees of freedom.
The theory predicts that up to corrections of order $1/m_{b,c}$, $m(B_s^0) -
m(B_d^0) = m(\Ds) - m(\Dd)$ \cite{mass-prediction}.
Recently, lattice QCD calculations have also given their predictions for the
meson mass spectrum \cite{mass-lattice1, mass-lattice2, mass-lattice3}.
%
%
By measuring the masses of mesons precisely, we narrow the range of parameters
and approximations that theoretical models use to make predictions.
For charm meson masses, a simultaneous fit \cite{PDG} of all measurements
including the mass difference between the \Ds and \Dd is used to compare
experimental measurements with theoretical predictions.
In this paper a measurement of the mass difference $m(\Ds)-m(\Dd)$ in the
decay channels $\Ds \rightarrow \phi \pi^+$ and $\Dd \rightarrow\phi\pi^+$ where
 $\phi \rightarrow K^+ K^- $ is presented \cite{notation}.
The advantage of measuring the mass difference in a common final decay state
is that many of the systematic uncertainties cancel.
Gathering the large sample of charmed mesons used in this analysis is done
using a novel displaced-track trigger, the Silicon Vertex Tracker (SVT) 
\cite{SVT-TDR}, which enables recognition of the decay of long-lived particles
early in the trigger system.

\section{The CDF II Detector and Data Set}

The data used for this analysis were collected with the upgraded Collider
Detector at Fermilab (CDF II) \cite{CDF-TDR} at the Tevatron $p \overline{p}$
collider.
The integrated luminosity is 11.6~$\rm pb^{-1}$ at $\sqrt{s} = 1.96~{\rm TeV}$,
taken during the period Feb - May 2002.
These are the first physics-quality data from the Run II program.

\subsection{The CDF II Detector}
\begin{figure}
\begin{center}
  \epsfig{file=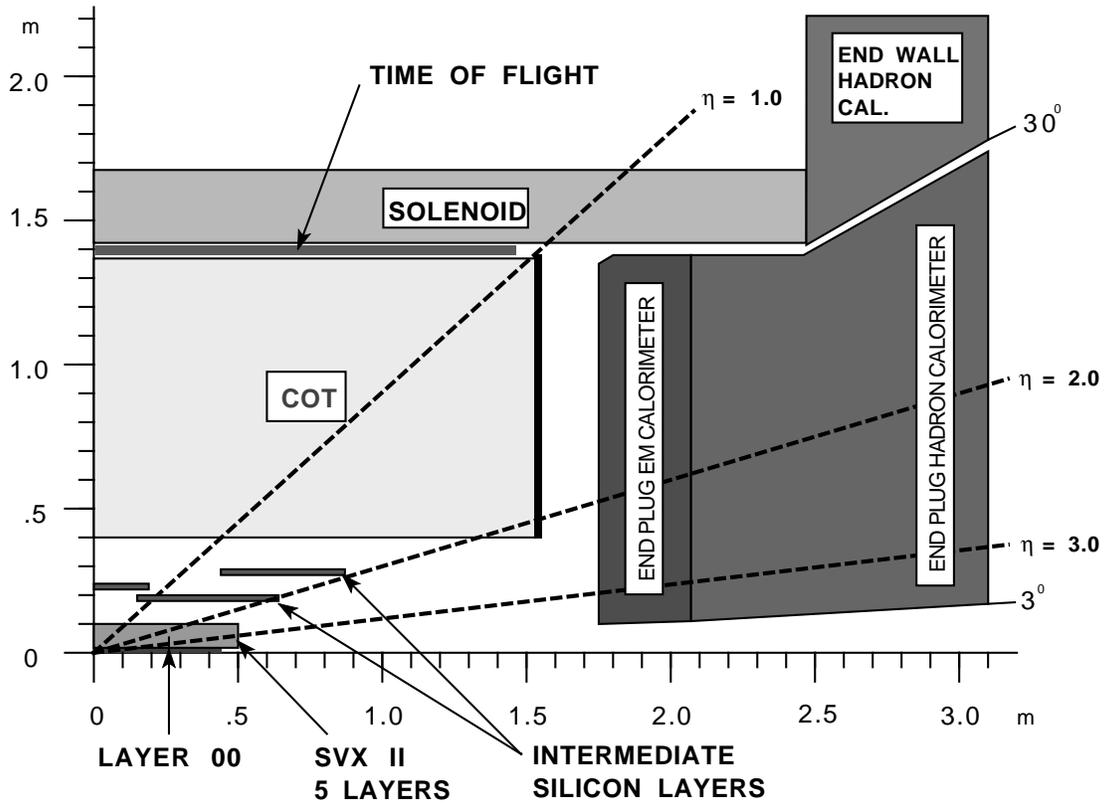,width=1.00\textwidth}
\end{center}
\caption{Quadrant view of the CDF II integrated tracking system. 
The Central Outer Tracker (COT) and silicon subdetectors form an
integrated tracking system.}
\label{fig:tracker}
\end{figure}

The CDF II detector is a major upgrade of the original CDF detector which last
took data in 1996.
The most important aspects of the upgraded detector for this analysis are the
new tracking system and the displaced track trigger.
CDF II, which is shown in Figure \ref{fig:tracker}, has an integrated central
tracking system immersed in a 1.4 T solenoidal magnetic field for the
measurement of charged-particle momenta.
The innermost tracking device is a silicon strip vertex detector, which
consists of three sub-detectors.
A single-sided layer of silicon sensors, called Layer 00 (L00) \cite{L00} is
installed directly onto the beryllium vacuum beam pipe, at a radius of 1.7 cm.
It is followed by five concentric layers of double-sided silicon sensors 
(SVXII) \cite{SVX} located at radii between 2.5 and 10.6 cm.
The Intermediate Silicon Layers (ISL) \cite{ISL} are the outermost silicon
sub-detector systems, consisting of one double-sided layer at a radius of 22 cm
in the central region, and two double-sided layers at radii 20 and 28 cm in
the forward regions.
Surrounding the silicon detector is the Central Outer Tracker (COT)
\cite{COT}, a 3.1 m long cylindrical open-cell drift chamber covering radii
from 40 to 137 cm.
The COT is segmented into eight super-layers, each consisting of planes of 12
sense wires.  
The super-layers alternate between axial wires and wires with a $\pm 2^\circ$
stereo angle, providing three-dimensional tracking. 
This provides up to 96 position measurements on a track passing through all eight
super-layers.
A charged particle traversing the tracking volume deposits charge on nearby
silicon micro-strips (clusters), and signals from the ionization trail in the
COT are recorded by the sense wires (hits).
Double-sided layers of silicon provide axial ($r$-$\varphi$) measurements of
cluster positions on one side and $z$ measurements via small-angle or
90-degree stereo information on the other. 
The L00 detector provides $r$-$\varphi$ measurements only.
COT information and SVXII $r$-$\varphi$ information from the SVXII detector are
used in this analysis.

\subsection{Tracking Parameters}

CDF II uses a cylindrical coordinate system ($r, \varphi, z$) with the origin at
the center of the detector and the $z$ axis along the nominal direction of the
proton beam. 
Tracks are fit to helical trajectories.
The plane perpendicular to the beam is referred to as the ``transverse plane'',
and the transverse momentum of the track is referred to as ${p_T}$.
%
%
%
In the transverse plane, the helix is parametrized with track curvature ($C$),
impact parameter ($d_0$) and azimuthal angle $\varphi_0$.
The projection of the track helix onto the transverse plane is a circle of 
radius $r$, and the absolute value of the track curvature is $|C| = 1/(2r)$.
The sign of the curvature matches the sign of the track charge.
The $d_0$ of a track is another signed variable; its absolute value
corresponds to the distance of closest approach of the track to the beam line.
The sign of $d_0$ is taken to be that of $\hat{ p}
\times \hat{d} \cdot \hat{z}$, where $\hat{p}$ and  $\hat{d}$ are unit
vectors in the direction of the particle trajectory and the direction of the
vector pointing from the primary interaction point to the point of closest
approach to the beam, respectively.
The angle $\varphi_0$ is the azimuthal angle of the particle trajectory at the
point of closest approach to the beam.
The two remaining parameters that uniquely define the helix in three
dimensions are the cotangent of the angle $\theta$ between the $z$ axis and
the momentum of the particle and $z_0$, the position along the $z$ axis at the
point of closest approach to the beam.
The two-dimensional decay length of a $D$ meson $L^D_{xy}$ is defined as 
\begin{equation}
 L^D_{xy} = \frac {\vec{X}_v \cdot \vec{P}_T^D }{ |\vec{P}_T^D |}
\end{equation}
where $\vec{P}_T^D$ is the transverse $D$ momentum and $\vec{X}_v$ is the
vector pointing from the primary interaction vertex to the $D$ meson decay
vertex. 
We use the average beam position as an estimate of the primary interaction
vertex.
This is calculated for each data acquisition run. 
The transverse intensity profile of the beam is roughly circular and can be
approximated by a Gaussian distribution with $\sigma \approx 35\,\mu{\rm m}$
\cite{SVT-PERF1, SVT-PERF2}.

\subsection{Trigger and Data Set}

CDF II has a three-level trigger system. The first two levels are implemented
with custom electronics, while the third is a software trigger based on a
version of the final reconstruction software optimized for speed.
The most important feature of the trigger system for this analysis is its
ability to recognize tracks and vertices displaced from the beam line.
A brief description of this part of the trigger system follows.
At Level 1 of the trigger, the COT provides information to the eXtremely Fast
Tracker (XFT) \cite{XFT} that identifies tracks with $p_T \geq 1.5~\GeVC$.
An event passes the Level 1 selection if the XFT finds a pair of tracks with
opposite charge, such that each has $p_T > 2.0~\GeVC$, the scalar sum of
transverse momenta $p_{T1} + p_{T2} > 5.5~\GeVC$ and angular difference
$\Delta \varphi_6 < 135 ^\circ$.
The angle $\varphi_6$ of a track is defined as the azimuthal angle of the track
momentum as measured in super-layer 6 of the COT, which corresponds to a
radius of 106 cm from the beam line.

At Level 2, the SVT combines XFT track information with SVXII information.
Tracks are refit using a linear algorithm, which provides improved $\varphi_0$
and $p_T$ measurements. 
The track impact parameter resolution is about $\rm 35~\mu m$ \cite{SVT-PERF1,
SVT-PERF2} for tracks with $p_T >~2~\GeVC$.
An event passes Level 2 selection if there is a track pair reconstructed in
the SVT such that each track has $p_T > 2.0~\GeVC$ and $100~{\rm \mu m}
< |d_0| < 1~{\rm mm}$.

At Level 3, the full three-dimensional track fit using COT information is
combined with SVT information.
The Level 2 requirements are confirmed with the improved track measurements.
The same tracks that passed the Level 1 selection have to pass the Level 2 and 
Level 3 requirements.
In addition, it is required that the vertex of the two trigger tracks has
$L_{xy} > 200\,\mu \rm m$.
The trigger requirements are optimized for selecting multi-body decays of
long lived charm and bottom mesons.  
The optimization is done using an unbiased trigger sample to estimate the
background rates and Monte Carlo simulated events to estimate the signal
rates.

Events gathered by the trigger system undergo final ``offline'' event
reconstruction with the best available tracking algorithms.
In the algorithm used for this measurement, the reconstruction begins with a
COT measurement of the track helix. 
This version of the track is extrapolated into the silicon tracker, starting
from the outermost layers and working inward.
Based on the uncertainties of the track parameters, a road is formed around the
extrapolated trajectory, and only silicon clusters found inside this road are
added to the track.
As clusters are added, the uncertainties on the track parameters are improved.
For this analysis, only the $r$-$\varphi$ information of the SVXII detector is
used.

\section{Momentum Scale Calibration}

The masses of the \Ds and \Dd mesons are measured from the momenta of their
decay daughters, therefore it is crucial to calibrate the momentum measurements
in the tracking volume.
The main effects that are of concern in this analysis are a proper accounting
of the energy loss in detector material and the calibration of the value of
the magnetic field ($B$).
Difficulties in accounting for energy loss in the tracking detectors come 
from an approximate model of the passive material.
Uncertainties of the magnetic field are determined directly from the data.
The momentum scale calibration for the tracking system is obtained by
studying a sample of $\sim$~55,000 $J/\psi \rightarrow \mu^+ \mu^-$ decays.
An incorrect accounting for material in the detector description causes the
reconstructed mass of the $J/\psi$ meson to depend on the its $p_T$.
Using an incorrect magnetic field value when converting track curvature into
momentum causes the mass of the $J/\psi$ meson to be shifted.
The calibration involves a two-step procedure. 
In the first step, the dependence of the $J/\psi$ mass on the transverse
momentum is eliminated by adding material to the tracking volume description.
After that, the magnetic field is calibrated by requiring that the reconstructed
$J/\psi \rightarrow \mu^+ \mu^-$ mass be equal to the world average.

\subsection{Procedure}

\begin{figure}
\begin{center}

  \epsfig{file=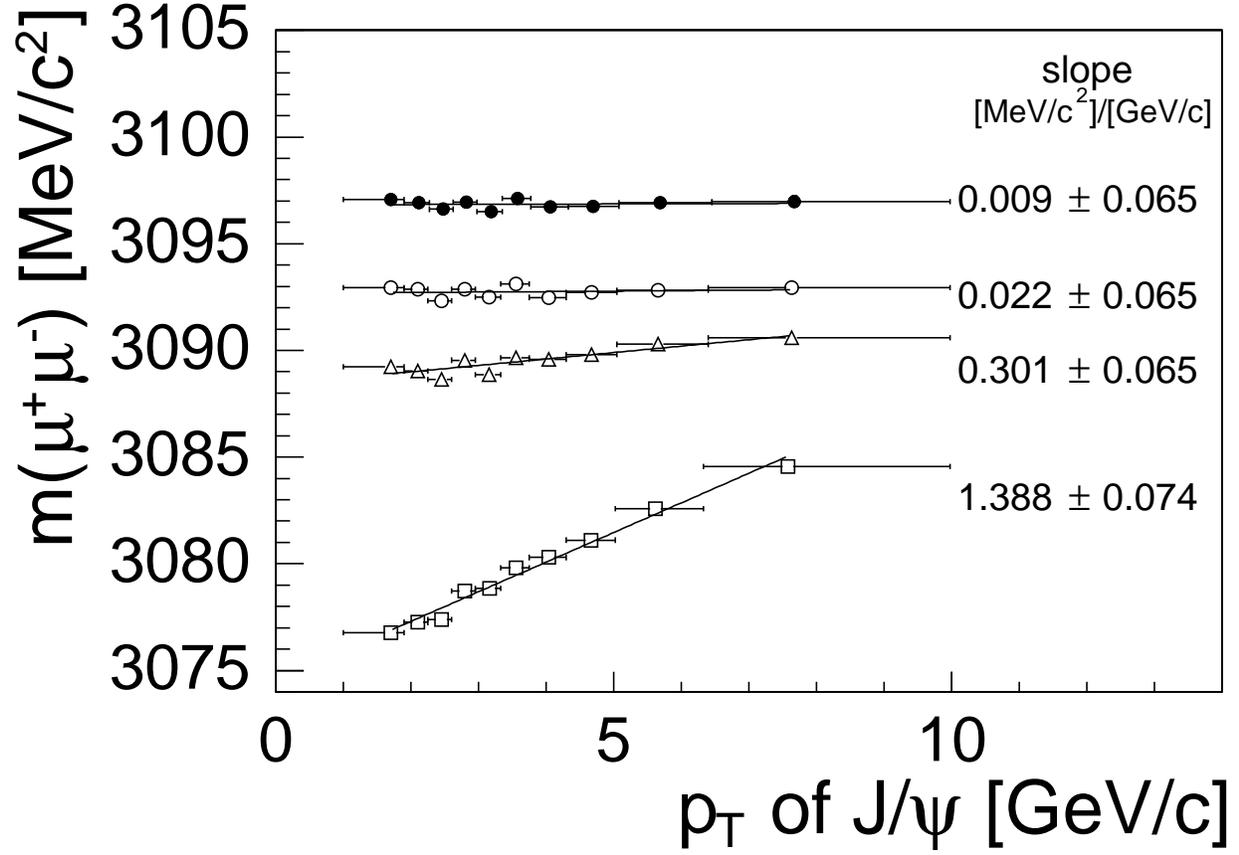,width=1.00\textwidth}
\end{center}
\caption{Dependence of the $J/\psi$ mass on the $p_T$ of the $J/\psi$. The open 
squares show the mass dependence for tracks with no energy loss
corrections. Open triangles show the result after applying the energy loss for
the material accounted for in the GEANT description of the detector. Open
circles account for the missing material modeled with the additional
layer. Filled circles show the effect of the $B$ field tuning in addition to
accounting for all the missing material.}
\label{fig:jpsiMassPt}
\end{figure}

The amount of passive material in the GEANT \cite{GEANT} description of the CDF
II silicon tracking volume is adjusted to eliminate the dependence of the
invariant mass of the $J/\psi$ candidates on their transverse momentum, as
demonstrated in Figure \ref{fig:jpsiMassPt}.
The missing material is modeled with a layer of uniform thickness located just
inside the inner shell of the COT; a layer of $0.56 \pm 0.10~{\rm g/cm^2}$
eliminates the dependence of the $J/\psi \rightarrow \mu^+\mu^-$ mass on its
$p_T$.
This additional layer corresponds to roughly $20\%$ of the total passive
material in the silicon tracking system.
Final state photon radiation causes a tail on the lower end of the $J/\psi$ mass
distribution, which distorts (compared with a Gaussian distribution) the shape
of the invariant mass distribution.
The corresponding bias is calculated in bins of $J/\psi$ momenta and is
taken into account when tuning the amount of passive material in the detector
description.

The magnetic field ($B$) is adjusted to bring the measured $J/\psi \rightarrow
\mu^+\mu^-$ mass to the world average value of $m(J/\psi) = 3096.87~{\rm \MeVCSq}$ 
\cite{PDG}.
The $B$ field is calibrated to a value of $ 1.41348 \pm 0.00027~{\rm T}$.
The precision of the tuning procedure is limited by the number of $J/\psi$
decays available for the calibration.

\subsection{Tests and Cross-Checks}

\begin{figure}
\begin{center}
  \epsfig{file=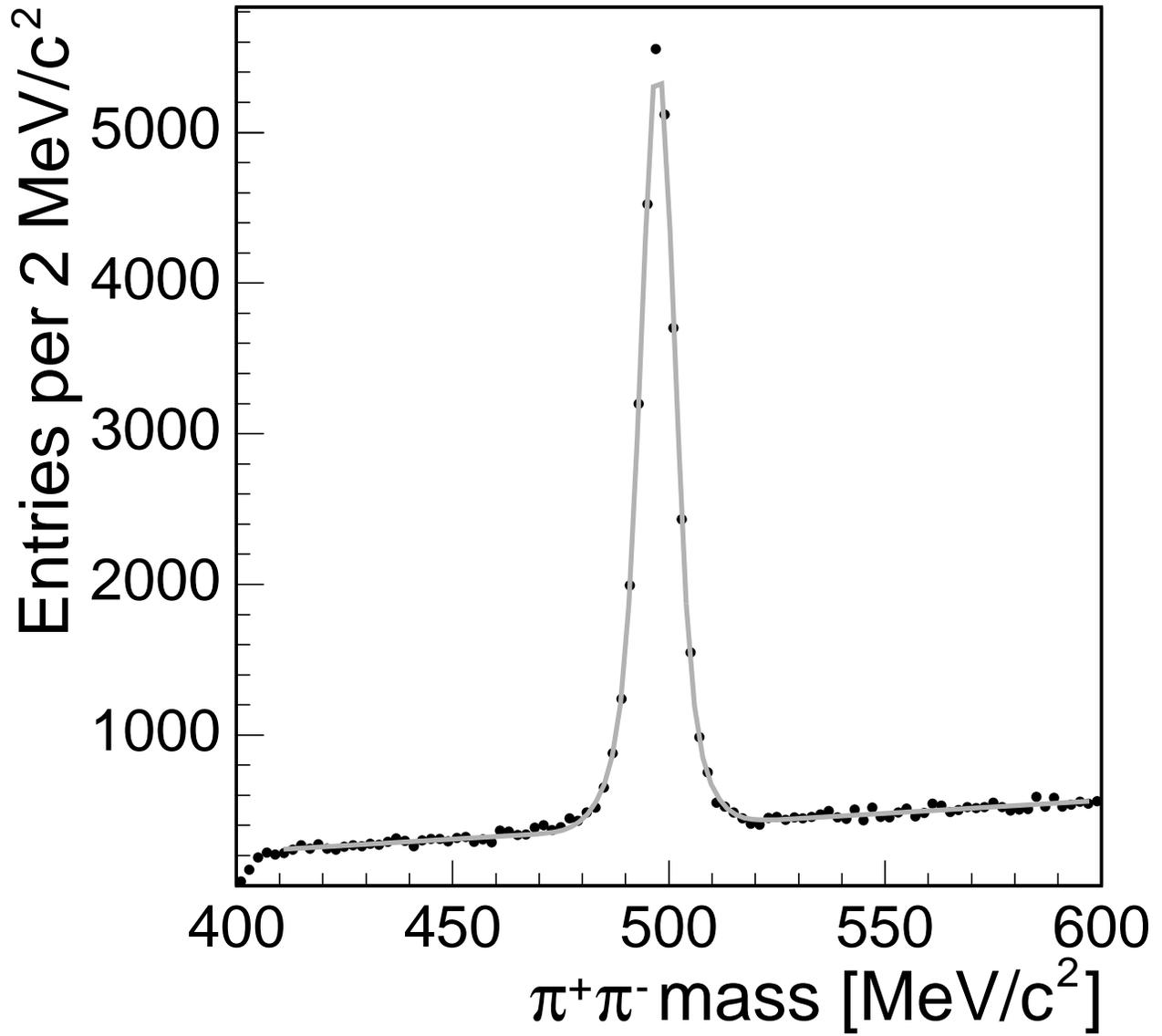,width=1.00\textwidth}
\end{center}
\caption{Measured $\pi^+ \pi^-$ mass distribution. A Gaussian distribution and a
  linear background are fitted to the mass spectrum.}
\label{fig:kshMass}
\end{figure}

\begin{figure}
\begin{center}
  \epsfig{file=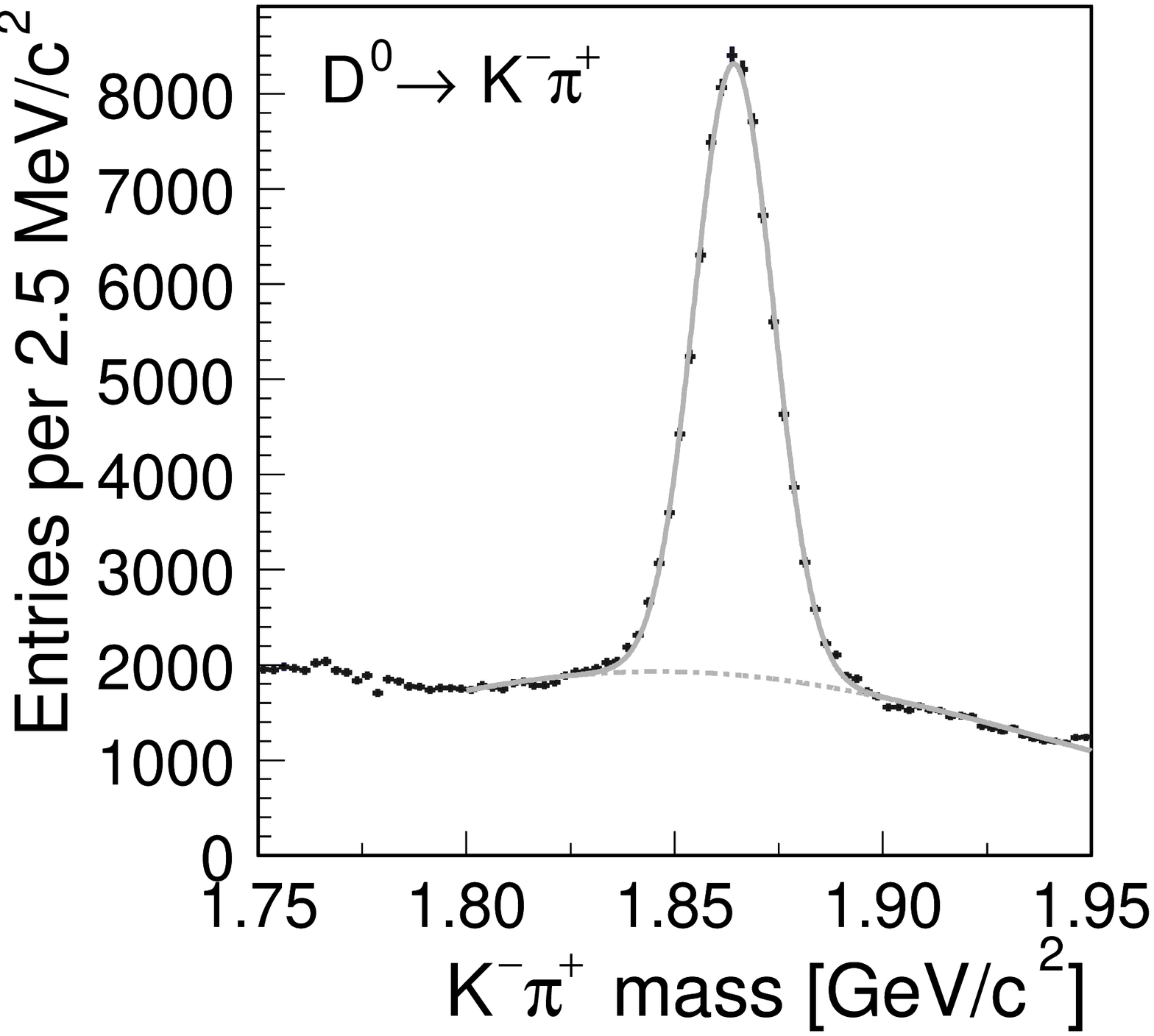,width=1.00\textwidth}
\end{center}
\caption{ The $K^-\pi^+$ mass distribution of the reconstructed $D^0$
candidates. A Gaussian distribution for the signal and a broad Gaussian
distribution for the background are fitted to the mass spectrum.}
\label{fig:d0MassPlot}
\end{figure}

\begin{figure}
\begin{center}
  \epsfig{file=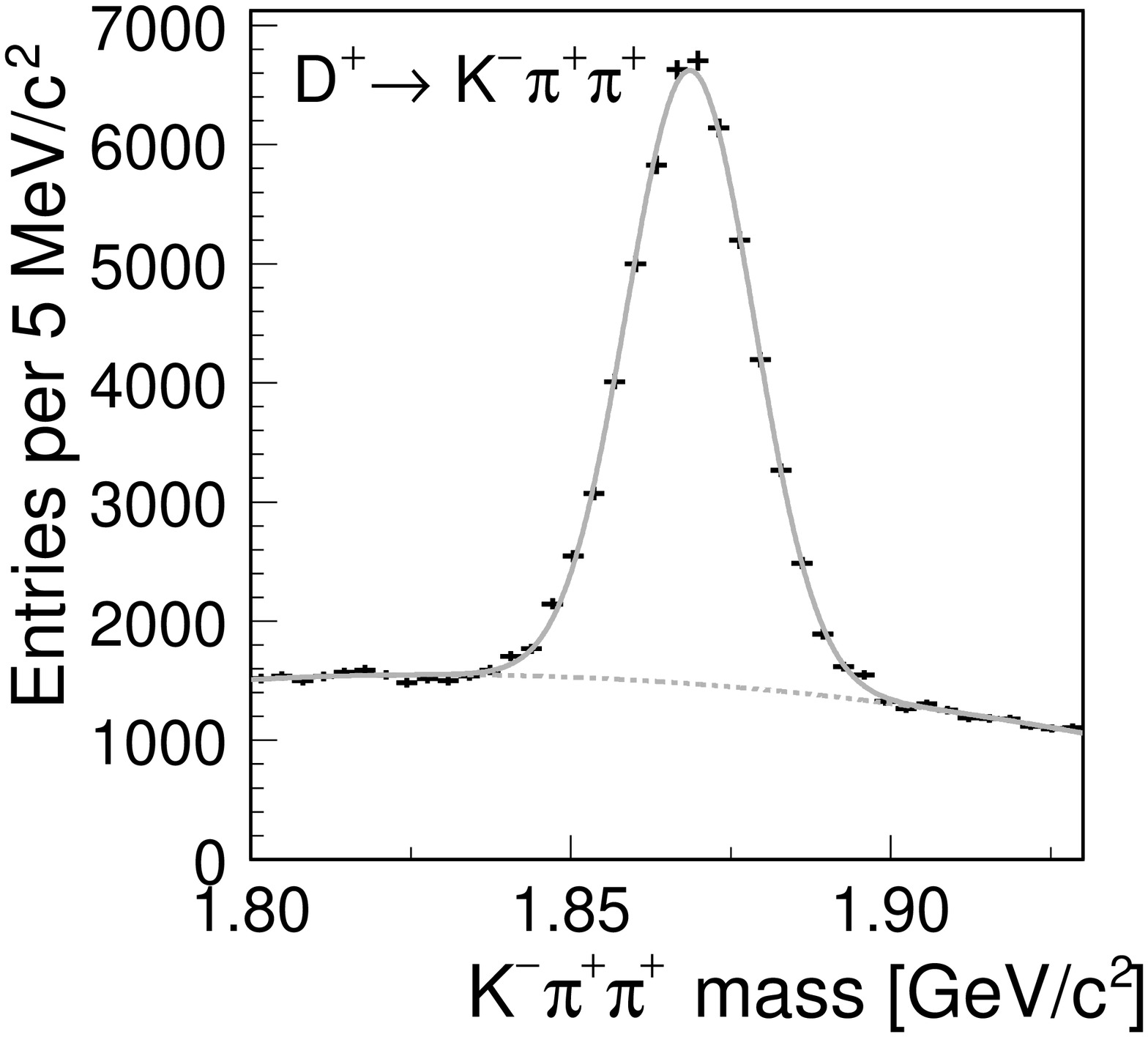,width=1.00\textwidth}
\end{center}
\caption{The $K^-\pi^+\pi^+$ mass distribution of the reconstructed $\Dd$ meson
candidates. A Gaussian distribution for the signal and a broad Gaussian
distribution for the background are fitted to the mass spectrum.}
\label{fig:d+Mass}
\end{figure}

\begin{figure}
\begin{center}
  \epsfig{file=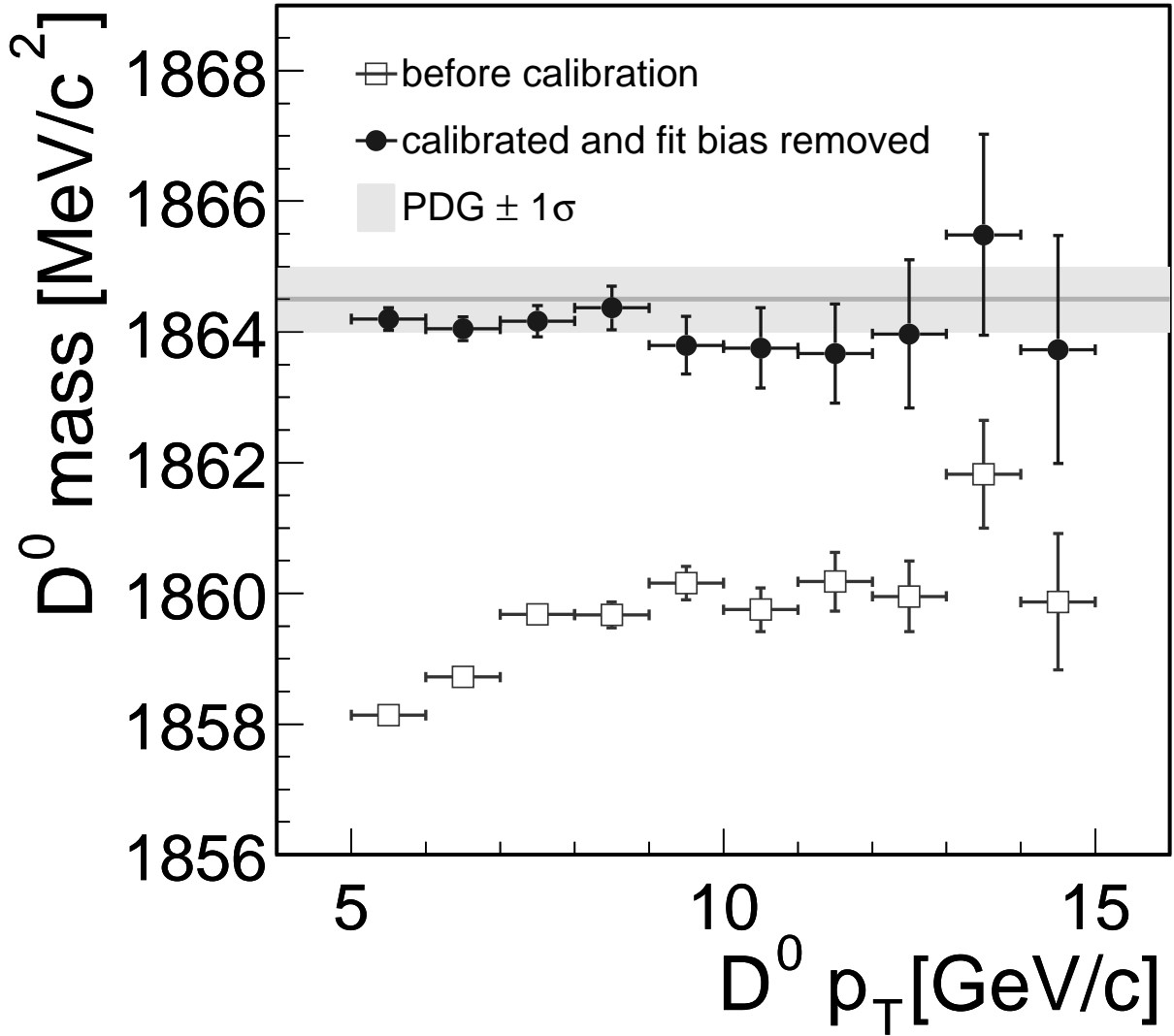,width=1.00\textwidth}
\end{center}
\caption{The dependence of the $D^0$ mass on its transverse momentum. The hollow
points show mass values before any corrections are applied; the solid points
show the dependence after the calibration (energy loss and $B$ field). The
systematic bias due to background modeling has been subtracted.}
\label{fig:d0MassPt}
\end{figure}

\begin{figure}
\begin{center}
  \epsfig{file=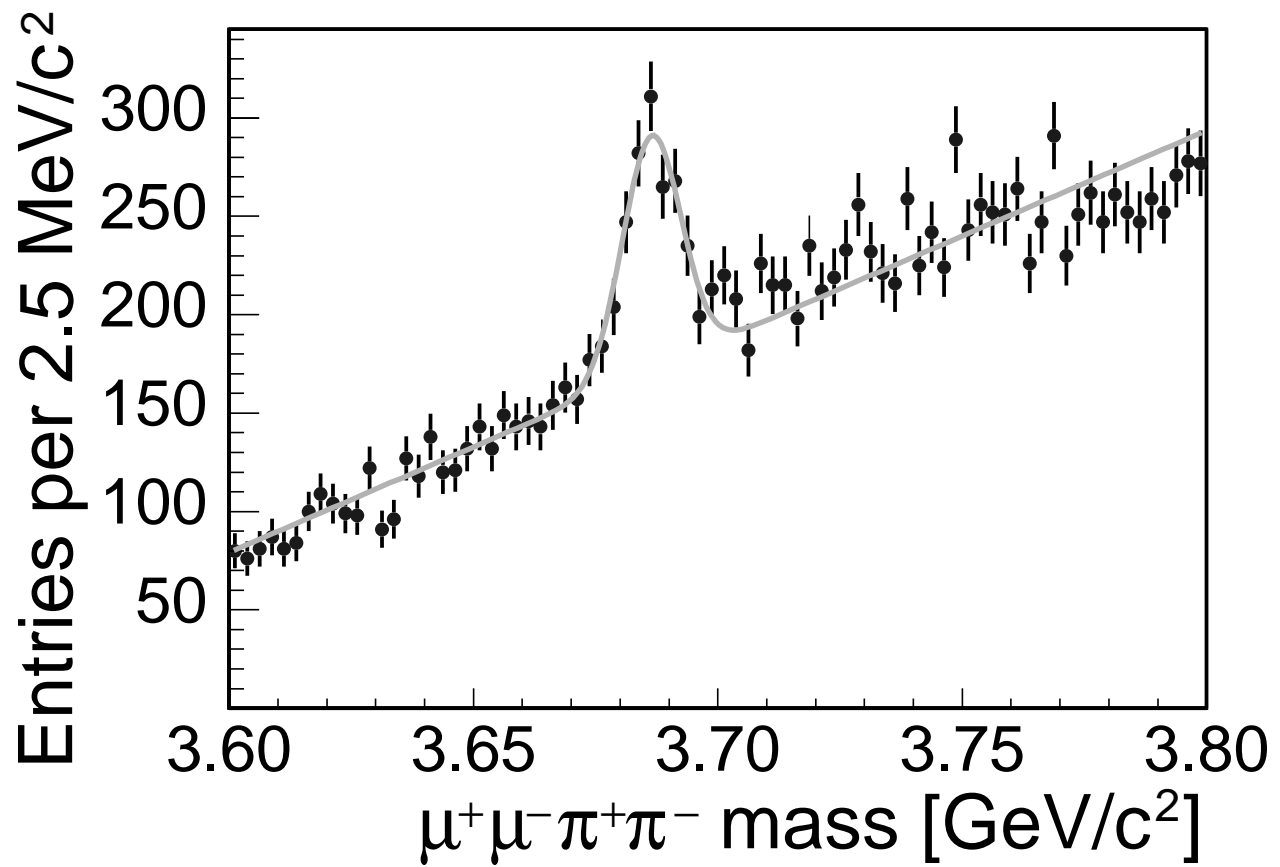,width=1.00\textwidth}
\end{center}
\caption{Measured $\mu^+\mu^- \pi^+\pi^-$ mass distribution for $\psi(2S)$
  candidates reconstructed in the $J/\psi~\pi^+\pi^-$ decay. A Gaussian
  distribution and a linear background are fitted to the measured spectrum.}
\label{fig:psiPrimeMass}
\end{figure}

\begin{figure}
\begin{center}
  \epsfig{file=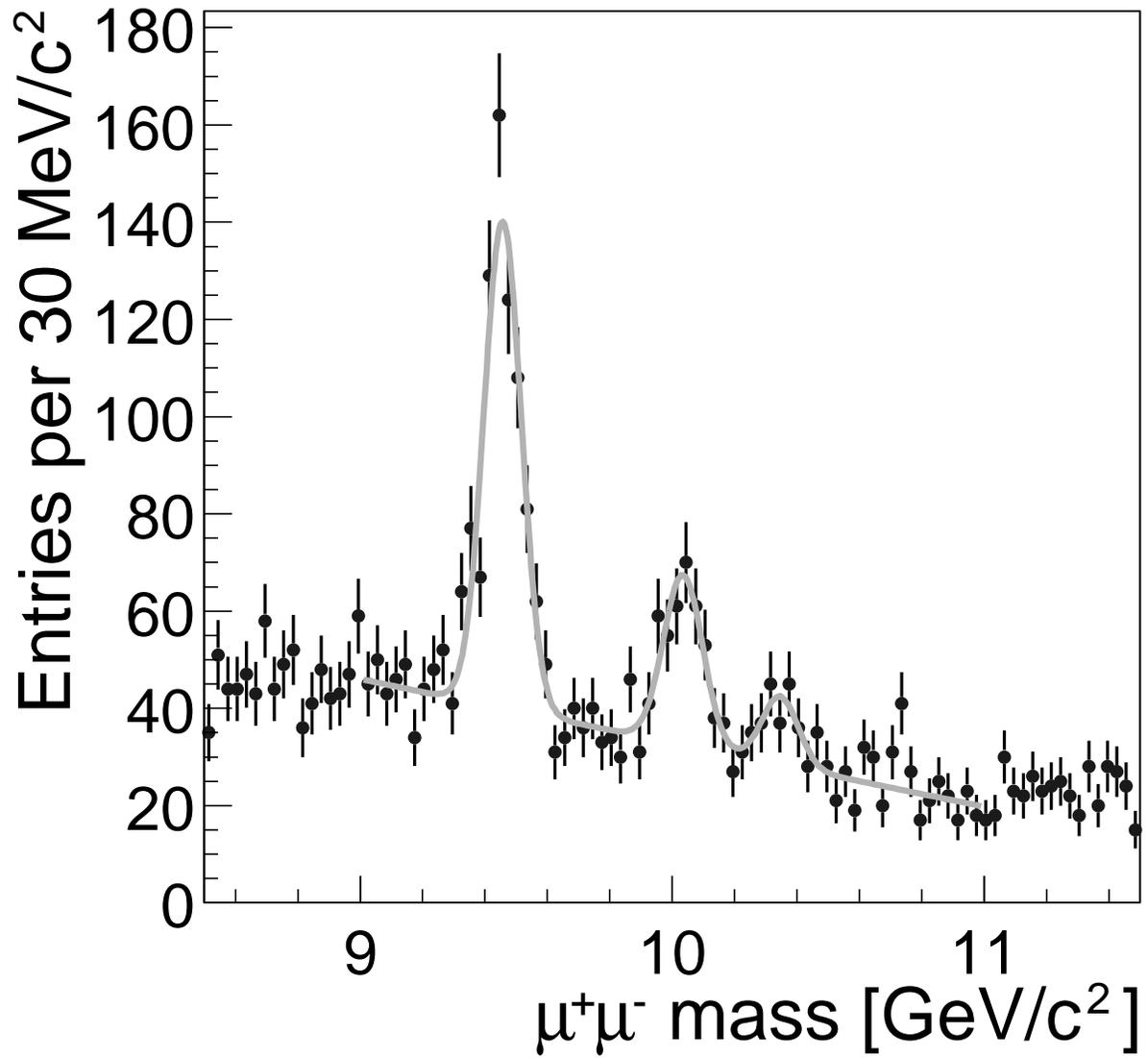,width=1.00\textwidth}
\end{center}
\caption{Measured $\Upsilon \rightarrow \mu^+ \mu^-$ mass distribution. Three
  Gaussian distributions and a linear background are fitted to the mass
  spectrum.}
\label{fig:upsilonMass}
\end{figure}

Several tests and cross-checks are performed to verify the calibration.
The $J/\psi$ invariant mass is checked for dependences on the $z$, $\varphi$ and
$\cot \theta$ coordinates of the decay in the detector. No significant
residual dependence is found after the calibration is applied.
The calibration method and parameters, the amount of missing passive material
and the magnetic field value, are also cross-checked with other meson decays
covering a range of invariant masses.
As a check in the low momentum range, $K_S^0 \rightarrow \pi^+ \pi^-$ decays are
studied.
The $\pi^+ \pi^-$ invariant mass distribution is presented in Figure \ref{fig:kshMass}. 
The $K_S^0$ decays are also studied for dependencies on the radial position of
the $K_S^0$ decay.
No significant dependence is found for radii several centimeters inside the
silicon detector.
The mass of the $K_S^0$ is checked for run-to-run variations. 
No significant dependence on the run number is found.
Cross-checks with high statistics, corresponding to several ten thousand signal
events, are done with samples of $D^0 \rightarrow K^- \pi^+$ and $D^+
\rightarrow K^- \pi^+ \pi^+$ decays presented in Figures \ref{fig:d0MassPlot}
and \ref{fig:d+Mass}, respectively.
The $D^0$ decays are also checked for mass dependence on the $p_T$ of the $D^0$.
Since no particle identification is used, there is a reflection peak in the
$D^0$ mass spectrum coming from the wrong assignment between kaon and pion
hypotheses that can not be removed.
The bias due to the reflection peak is estimated using a parametric simulation
for every $p_T$ bin separately and taken into account in Figure
\ref{fig:d0MassPt}.
The $\psi(2S) \rightarrow J/\psi \pi^+ \pi^-$ decays are also reconstructed and
the mass distribution is shown in Figure \ref{fig:psiPrimeMass}.
Finally, a check in the region of higher momenta is done with $\Upsilon
\rightarrow \mu^+ \mu^-$ decays, presented in Figure \ref{fig:upsilonMass}.
The reconstructed masses are compared to the world average values \cite{PDG}
in Table \ref{tab:massCompendium}.
We conclude that the calibration procedure described above accounts well for
the energy loss in the silicon tracking volume, and applies to a range of
reconstructed invariant masses.
The calibration parameters quoted above are used when reconstructing the
invariant mass of the $\Ds$ and $\Dd \rightarrow \phi \pi$ decays.

\begin{figure}
\begin{center}
  \epsfig{file=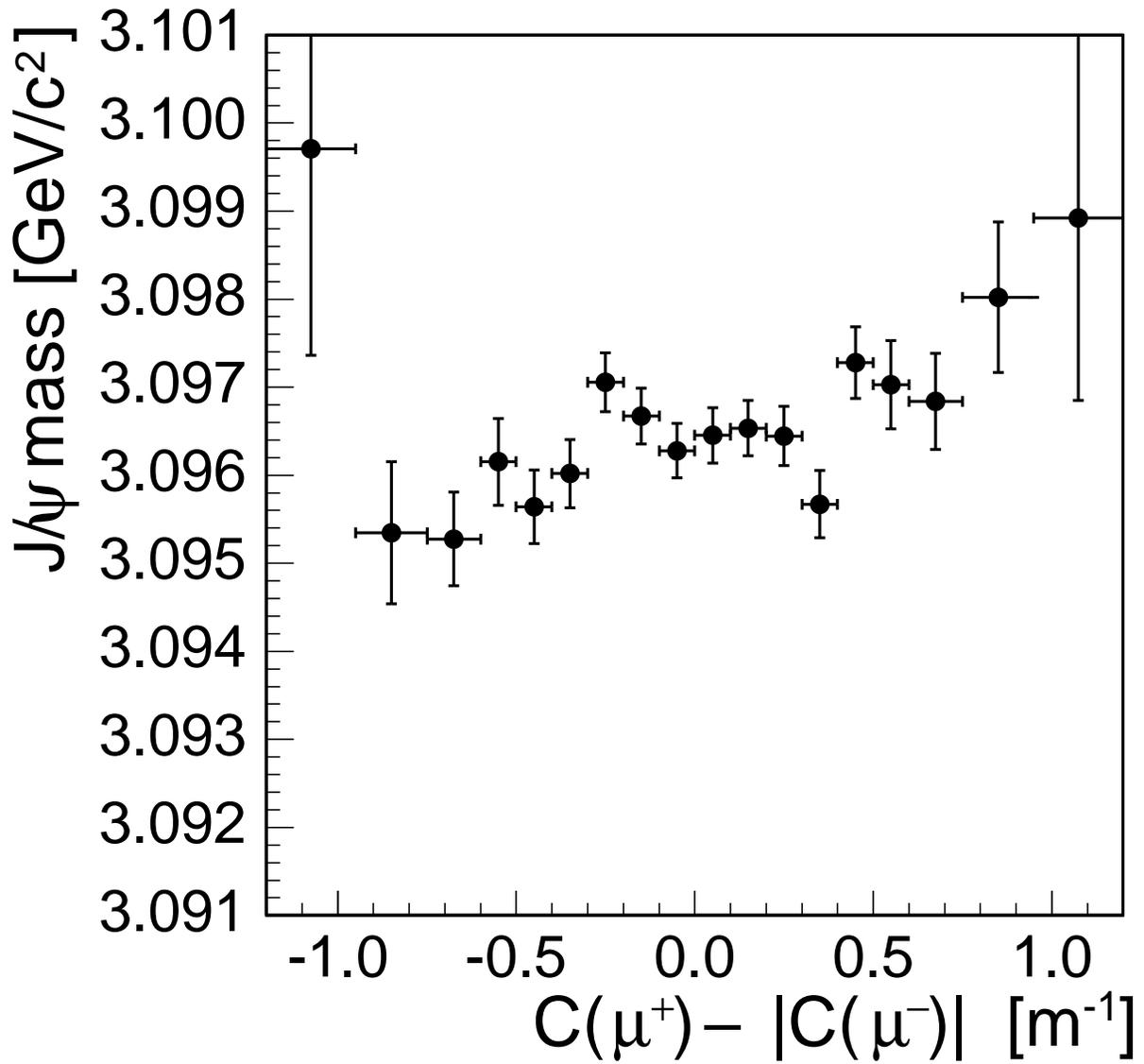, width=1.00\textwidth}
\end{center}
\caption{Dependence of the $J/\psi$ mass on the difference of the
absolute values of the curvature ($C$) of the positive and negative muon. This
distribution shows a small charge dependent effect that are not corrected for
in the calibration. }
\label{fig:falseCurvature}
\end{figure}

One effect is found that is not completely corrected by the calibration.
The distribution of the invariant mass of the $J/\psi$ as a function of the
curvature difference between the two muons shows a slope, as seen in Figure
\ref{fig:falseCurvature}.
This dependency indicates charge specific effects in the tracker, referred to
as ``false curvature''.
It also manifests itself in a difference in mass of the charge conjugates of
the same meson.
Misalignments in the COT, relative alignment of the COT to the silicon
tracker, tilted wire planes and discrepancies between the COT axis and the
magnetic field axis can cause such charge dependent false curvature effects.
Parametrized corrections applied to track parameters improve  the
distribution shown in Figure \ref{fig:falseCurvature}. 
The charge asymmetry of the mass of charged mesons is not eliminated by these
corrections.
We do not correct for false curvature effects in the calibration procedure, but
instead estimate the systematic uncertainty arising from the observed asymmetry.

\boldmath
\section{$\Ds$ and $\Dd$ Selection}
\unboldmath

The \Ds and~\Dd mesons are selected using offline reconstructed tracks through
their decays to $\phi\pi^+$ followed by the subsequent decay $\phi\rightarrow
K^+ K^-$.
To ensure good track quality, the tracks are required to have hits in $\geq
20$ COT stereo layers, $\geq 20$ axial layers, $\geq 3$ silicon $r$-$\varphi$
clusters and $p_T > 400~\MeV/\!c$.
No particle identification is used in this analysis, and all mass assignments
consistent with the assumed decay are attempted.

The $\phi$ candidates are selected by requiring two charged tracks, assumed to
be kaons, which have opposite charge. 
The invariant mass of the track pair is required to be within $10~\MeVCSq$ of
the world average $\phi$ mass.
The detector resolution of the $\phi$ mass is approximately $4~\MeVCSq$.
A third track, assumed to be a pion, is added to the $\phi$ candidate.
To avoid using tracks from different interaction vertices, the separation
along the beam line of all three tracks, the two kaon candidates and the pion
candidate, is required to be $<4\,\rm cm$.
Any two of these three tracks satisfy trigger-like criteria using offline
quantities: opposite charge, $p_T > 2.0~\GeVC$, and
$120\,\mu{\rm m} < |d_0| < 1\,{\rm mm} $.
The third track is only required to have $|d_0| < 2~\mathrm{mm}$. 
No further requirements are placed on this track.

All three tracks are constrained to a common vertex in 3 dimensions.
To ensure quality of the vertices, the $\chi^2$ of the vertex
in the transverse plane satisfies $\chi^2(r,\varphi) < 7$.
The displaced track trigger preferentially accepts events with two-track
vertices displaced from the primary interaction point by a few hundred
microns.
Adding a third track from the primary interaction pulls the three-track vertex
toward the beamline, and the resulting $L_{xy}$ of the three track vertex is
much smaller.
To eliminate these background candidates, the $L_{xy}$ of the three track
vertex is required to be larger than $500~\um$.

\begin{figure}
\begin{center}
  \epsfig{file=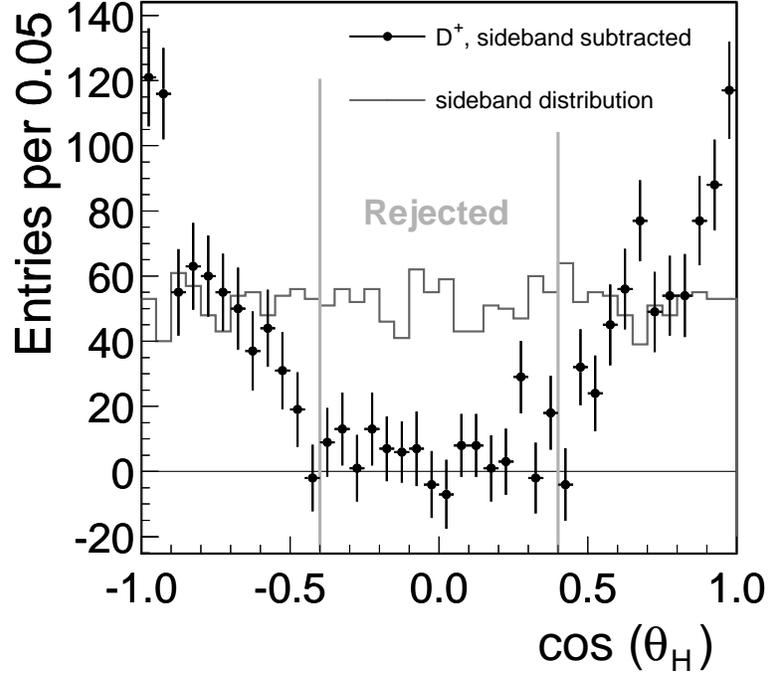,width=0.63\textwidth}\\
  \epsfig{file=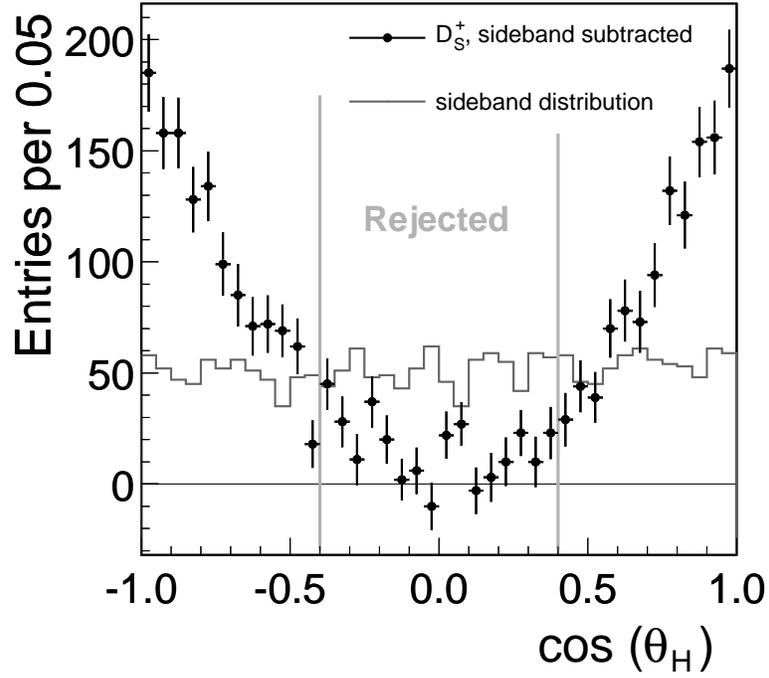,width=0.63\textwidth}
\end{center}
\caption{Sideband subtracted and sideband distributions of the cosine of the
helicity angle of the \Dd candidates (left) and \Ds candidates (right).
Candidates with $|cos(\theta_H)| < 0.4$ are rejected in the selection.}
\label{fig:helicity}
\end{figure}

The helicity angle ($\theta_H$) is defined as the angle between the $\phi$
flight direction and the direction of the kaon momentum measured in the $\phi$
rest frame.
The $\phi$ is polarized in this decay channel, so the helicity angle is
expected to follow a $\cos^2\theta_H$ distribution for the signal, and a flat
distribution for the background.
Using sideband subtraction, we verify that the other selection requirements
do not distort the shapes of these distributions, as demonstrated in Figure
\ref{fig:helicity}.
The helicity angle is required to satisfy $|\cos(\theta_H)| > 0.4$.

The requirements on the fit $\chi^2(r,\varphi)$, $L_{xy}$ and $|\cos(\theta_H)|$
have similar efficiencies.
Individually, each requirement is $90-95\%$ efficient for the signal
candidates, and rejects $40-50\%$ of the background.
It is unlikely to find two real $\Ds / \Dd \rightarrow \phi \pi^+$ decays in
the same event.
If multiple candidates are found in an event, only the candidate with the
highest $|\cos(\theta_H)|$ is considered.
This procedure rejects another $9\%$ of the underlying background.
%

\section{Mass Fitting and Systematic Uncertainties}

\begin{figure}
\begin{center}
  \epsfig{file=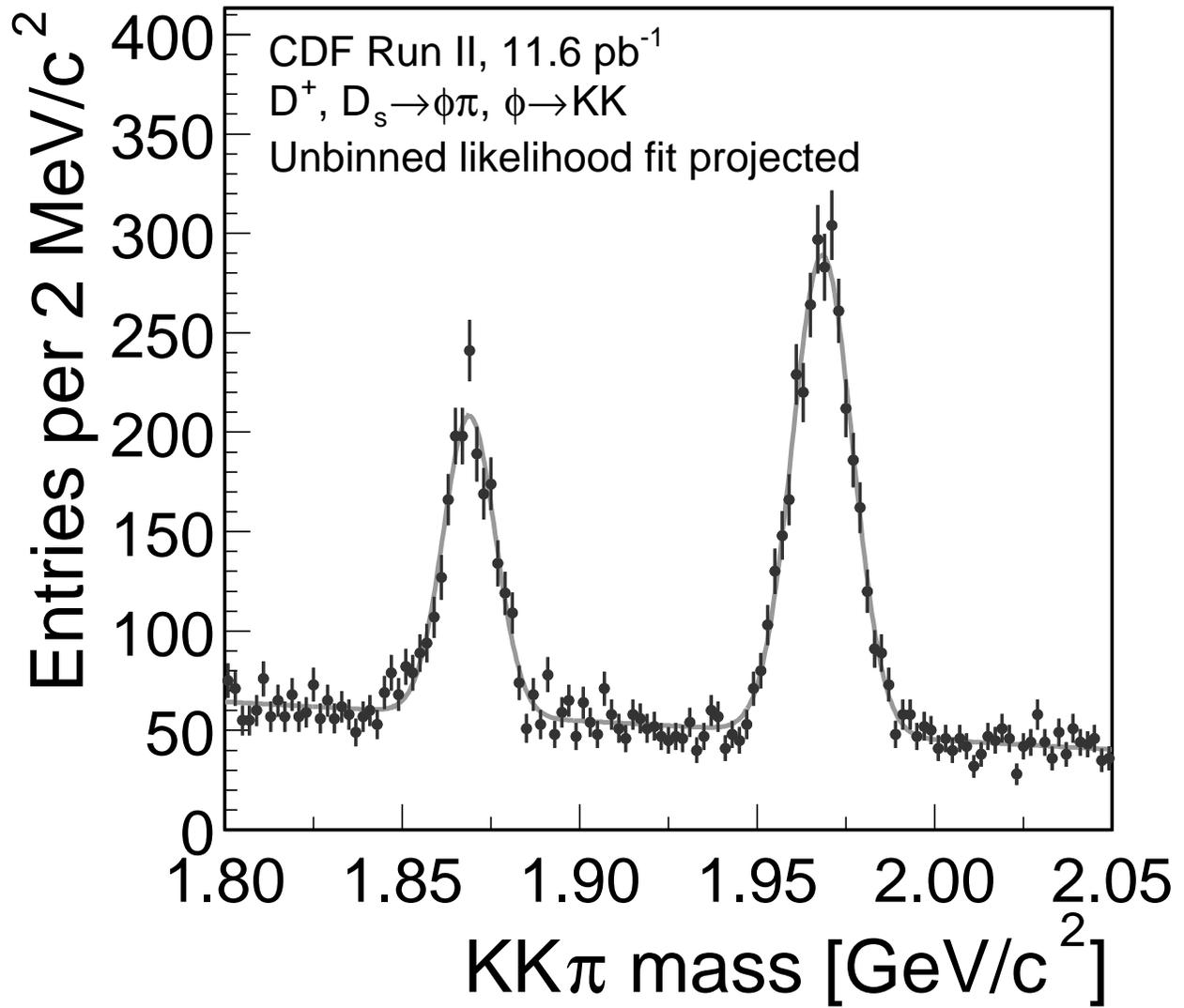,width=1.00\textwidth}
\end{center}
\caption{Measured $K^+K^-\pi^+$ mass distribution compared to the unbinned
likelihood fit. }
\label{fig:baseProjection}
\end{figure}

The invariant mass distribution of the $K^+ K^- \pi^+$ system is fit to two
Gaussian distributions and a linear background.
An unbinned maximum likelihood fit is used in which the widths of both Gaussian
distributions, the mass of the \Ds and the $m(\Ds) - m(\Dd)$ mass difference are
allowed to float independently.
Studies on both data and Monte Carlo simulation show that a linear dependence
on mass is a good description of the background.
Figure \ref{fig:baseProjection} shows the likelihood fit superimposed onto the
invariant mass spectrum.
The $\chi^2$ of the comparison of the likelihood fit to the measured mass
spectrum is 127 for 118 degrees of freedom, and corresponds to a $\chi^2$
probability of 27\%.
The complete list of fit parameters can be found in Table \ref{tab:fit}, and
the fit result yields:
\begin{equation}
 m(\Ds) - m(\Dd) = \rm 99.41 \pm 0.38~(stat) ~\MeVCSq.
\end{equation}

\begin{figure}
\begin{center}
  \epsfig{file=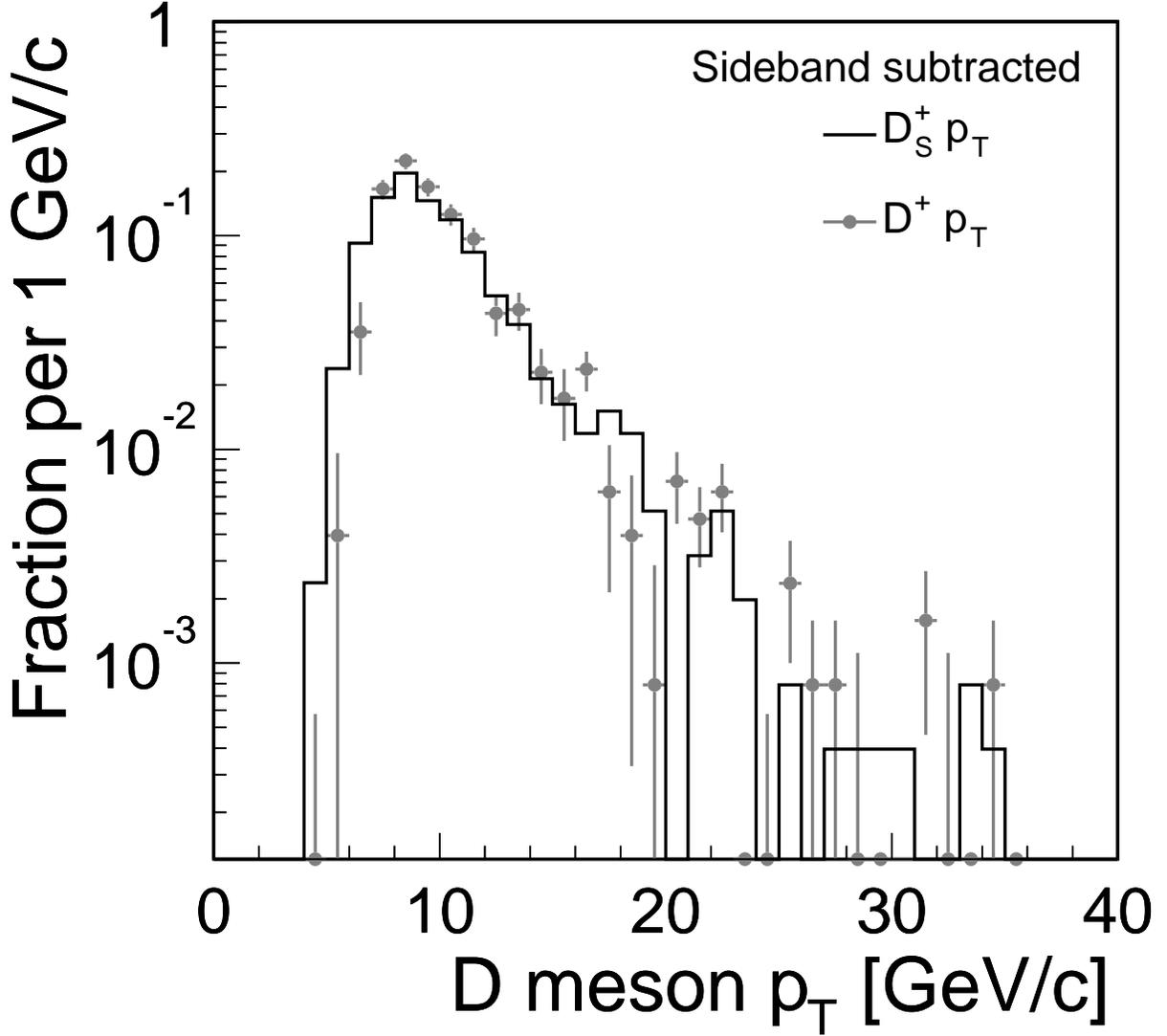,width=1.00\textwidth}
\end{center}
\caption{Sideband subtracted distributions of the $p_T$ of the $\Ds$ candidates
  (solid) and $\Dd$ candidates (dots). Both distributions are normalized such
  that the sum of the bins add up to one.}
\label{fig:comparePt}
\end{figure}

The two charmed mesons are produced either directly in the $p\overline{p}$
collision, or they are products of a $B$ meson decay.
The trigger preferentially selects mesons with large displacements of the
decay vertex from the primary interaction point.
Since the \Ds and \Dd mesons have different lifetimes, the fraction of
directly produced \DsNoSpc / \Dd mesons to those coming from $B$ meson decays
is also different.
Therefore, the momentum spectra of the two signals may differ, causing
differences in the final state kinematics.
This kinematic difference can produce a systematic shift in the measurement 
of the mass difference.
Figure \ref{fig:comparePt} shows a comparison of the $p_T$ distributions of
the $\Ds$ (solid line) and the $\Dd$ (dotted line).
The spectra are very similar, and we expect small systematic uncertainties.

\subsection{Discussion of Systematic Uncertainties}

The systematic uncertainties are summarized in Table~\ref{tab:systematics},
and will now be discussed in order of decreasing size.
The largest single systematic uncertainty comes from fitting.
To estimate the systematic uncertainties due to background modeling, the
results of fits with different background models are compared.
One model used in this comparison is a linear combination of orthogonal
polynomials.
Another model consists of two piecewise linear functions that meet at a point, 
which is varied between the $\Dd$ and $\Ds$ mass distributions.
A systematic uncertainty of $0.08~\MeVCSq$ on the mass difference is
assigned based on the variation of the fit result when these different models
are used.
The systematic effect due to signal modeling is studied by excluding regions
of the $\Ds$ and $\Dd$ signals from the fit.
In this case, a fraction of the variation of the fit result is caused by
changing the statistics of the sample used.
This contribution is estimated by comparing statistical uncertainties of the fit
result with regions excluded to that of the fit result with no modification.
After estimating the statistical contribution of the variation of the fit
result, the systematic uncertainty due to signal modeling is estimated to be
$0.12~\MeVCSq$.
These two systematic uncertainties are added in quadrature and a systematic
uncertainty of $0.14~\MeVCSq$ due to fitting is obtained.
This is the largest single systematic uncertainty.

To estimate the systematic uncertainty introduced by sample selection
requirements, the requirements on $\chi^2(r,\varphi)$, $L_{xy}$, $\cos
\theta_H$, and duplicate rejection are individually varied.
Fit results were compared to estimate systematic effects for individual
selection requirements.
A fraction of the variation in the fit result is caused by statistical effects
due to changing the sample composition when the selection requirements change.
As before, the statistical contribution to the fit result variation is estimated
from the change in the statistical uncertainty of the fit result.
The only relevant selection requirement which exhibits a statistically
significant effect is the cut on the $\chi^2(r,\varphi)$ variable.
This variation of the mass difference is traced to an enhanced background around
the $D^+$ mass for small values of the $\chi^2(r,\varphi)$ variable.
The effect is estimated to cause a systematic uncertainty of $0.11~\MeVCSq$.


The systematic uncertainty due to the momentum scale determination is
estimated by analyzing a kinematically similar decay.
A GEANT study is done to determine how the uncertainty on the mass difference
measurement would scale with the absolute uncertainty on the $\Dd \rightarrow K
\pi \pi$ mass due to momentum scale variations, and shows that the uncertainty
on the mass difference corresponds roughly to 11\% of the absolute uncertainty
on the \Dd mass.
The world average mass of the $\Dd$ meson $m(\Dd) = 1869.4 \pm 0.5~\MeVCSq$ is
compared to our measurement of $m(\Dd) = 1868.65 \pm 0.07~\MeVCSq$ obtained in
a sample of $\Dd\rightarrow K^- \pi^+ \pi^+$ decays, using the same
calibration procedure.
To determine the absolute uncertainty of the momentum scale, the uncertainty of
the world average ($0.5~\MeVCSq$), the statistical uncertainty of our
measurement ($0.07~\MeVCSq$) and the difference between the two measurements
(0.75~\MeVCSq) are added in quadrature.
The sum in quadrature is then scaled by the factor obtained in the Monte Carlo
study, and the systematic uncertainty of the momentum scale determination is
estimated to be $\rm 0.10~\MeVCSq$.

The mass difference is also sensitive to detector effects that are not
corrected for by our calibration, namely false curvature effects.
These effects are expected to cancel in the measurement of the mass
difference.
As explained in the calibration section, empirical corrections of the track
curvature do not completely eliminate the asymmetry of charge conjugate 
states.
By comparing fit results with and without these empirical corrections,
the systematic effect of uncorrected tracking effects is estimated to be
$0.06~\MeVCSq$.

The accuracy of the momentum scale calibration is limited by the size of the
$J/\psi$ sample.
The systematic uncertainty on the mass difference from this limitation is estimated
by individually varying the amount of material and the magnitude of the
magnetic field by their statistical precisions.
The two systematic effects are added in quadrature to obtain a systematic
uncertainty of $0.03~\MeVCSq$ due to the calibration procedure.

Finally, an explicit check is done for a systematic uncertainty caused by the
difference in $p_T$ spectra of the \Ds and \Dd shown in Figure \ref{fig:comparePt}.
The events were re-weighted in the fit to make the spectra identical and the
systematic effect on the mass difference is found to be negligible.

The total systematic uncertainty of the measurement is estimated by combining
the above systematic uncertainties in quadrature, and is found to be
$0.21~\MeVCSq$.

\section{Summary}

The difference between the mass of the $\Ds$ meson and the $\Dd$ meson is
measured using 11.6 $\rm pb^{-1}$ of data collected by CDF II and is found to be
\begin{center}
  $m(\Ds) - m(\Dd) = \rm 99.41 \pm 0.38 (stat.) \pm 0.21 (syst.)
  ~\MeVCSq$.
\end{center}
The result is in agreement with the current world average~\cite{PDG} and the
most recent Babar publication of
$\left(98.4\pm0.1\pm0.3\right)~\MeVCSq$~\cite{BaBar}, with a comparable
uncertainty.

\section*{Acknowledgments}
We thank the Fermilab staff and the technical staffs of the participating
institutions for their vital contributions. We especially acknowledge the
contributions of the members of the Fermilab beams division. This work was
supported by the U.S. Department of Energy and National Science Foundation;
the Italian Istituto Nazionale di Fisica Nucleare; the Ministry of Education,
Culture, Sports, Science and Technology of Japan; the Natural Sciences and
Engineering Research Council of Canada; the National Science Council of the
Republic of China; the Swiss National Science Foundation; the A.P. Sloan
Foundation; the Bundesministerium fuer Bildung und Forschung, Germany; the
Korean Science and Engineering Foundation and the Korean Research Foundation;
the Particle Physics and Astronomy Research Council and the Royal Society, UK;
the Russian Foundation for Basic Research; and the Comision Interministerial
de Ciencia y Tecnologia, Spain.

\clearpage

\bibliography{delMass-prd-1}

\clearpage

\begin{table}
\begin{center}
\begin{tabular}{c @{$\rightarrow\,$}l r@{$\,\pm\,$}l r@{$\,\pm\,$}l }
\multicolumn{2}{l}{Decay} & \multicolumn{2}{c}{Mass $\rm [\MeVCSq]$} & \multicolumn{2}{c}{PDG $\rm[\MeVCSq]$} \\
\hline
$K_S^0 $ & $\pi^+\pi^-$     &  $ 497.36$ & $ 0.04$ & $497.672$ & $ 0.031$  \\
$\Upsilon$ & $\mu^+ \mu^-$  &  $9461   $ & $ 5   $ & $9460.30$ & $ 0.26 $  \\
$D^0$ & $K^- \pi^+$         &  $1864.15$ & $ 0.10$ & $1864.5 $ & $ 0.5  $  \\
$D^+$ & $K^- \pi^+ \pi^+$   &  $1868.65$ & $ 0.07$ & $1869.4 $ & $ 0.5  $  \\
$\psi(2S)$ & $J/\psi\pi^+\pi^-$&  $3686.43$ & $ 0.54$ & $3685.96$ & $ 0.09 $  \\
\hline
\hline
\end{tabular}
\end{center}
\caption{Table comparing measured masses of mesons reconstructed using the
described calibration parameters and corresponding PDG averages. Uncertainties
on reconstructed masses are statistical only.}
\label{tab:massCompendium}
\end{table}

\begin{table}
\begin{center}
{
 \begin{tabular}{ l r @{$\,\pm\,$} l c }
  Parameter                 &  \multicolumn{2}{c}{Value} \\
  \hline                                      
  $\delta m$    \hfill [$\MeVCSq$]        &   99.41 & 0.38  \\
  $m(D_s)$      \hfill [$\MeVCSq$]        & 1968.4  & 0.3   \\
  $\sigma(D_s)$ \hfill [$\MeVCSq$]        &    8.4  & 0.2   \\
  $\sigma(D^+)$ \hfill [$\MeVCSq$]        &    7.3  & 0.3   \\
  $f(D_s)$                                &   0.65  & 0.01  \\
  $f(D^+ + D_s)$                          &   0.37  & 0.01  \\
  background slope \hfill[1/$\GeVCSq$]    &   -7.3  & 0.7   \\
  \hline
  $\chi^2$/NDF                            & \multicolumn{2}{l}{126.7/118 (27.9~\%)} \\  
  \hline \hline
  \end{tabular}
}
\end{center}

\caption{Table of likelihood fit parameter results corresponding to Figure
\ref{fig:baseProjection}.  The $\chi^2$, number of degrees of freedom (NDF) and
corresponding probability are also listed. The parameters are the mass
difference ($\delta m$), the mass of the $\Ds$ meson, the mass resolutions
($\sigma(D_s)$, $\sigma(D^+)$), the fraction of signal events ($f(D_s)$ , $f(D^+
+ D_s)$) and the slope of the background.}
\label{tab:fit}
\end{table}

\begin{table}
\begin{center}
\begin{tabular}{lc}
Effect & Syst.$ [\rm \MeVCSq]$ \\
\hline
fitting (signal + background)  & 0.14 \\ 
\hline
             event selection   & 0.11 \\
\hline
              momentum scale   & 0.10 \\ 
             tracker effects   & 0.06 \\
        calibration procedure  & 0.03 \\
\hline
Total                          & 0.21 \\
\hline
\hline
\end{tabular}
\end{center}
\caption{ Table of systematic uncertainty estimates for the mass difference. The
total uncertainty is the quadratic sum of the individual uncertainties.}
\label{tab:systematics}
\end{table}

\clearpage

\clearpage

\clearpage

\clearpage

\clearpage

\clearpage

\clearpage

\clearpage

\clearpage

\clearpage

\clearpage

\clearpage

\end{document}